\documentclass[11pt]{article}
\usepackage{moriond,epsfig}

\def\Journal#1#2#3#4{{#1} {\bf #2}, #3 (#4)}

\def\ApJ{{\em ApJ}}
\def\MNRAS{{\em MNRAS}}

\def\be{\begin{equation}}
\def\ee{\end{equation}}
\def\bea{\begin{eqnarray}}
\def\eea{\end{eqnarray}}

\begin{document}
\vspace*{4cm}
\title{THE DEPENDENCE OF OBSERVED GAMMA-RAY PULSAR \\ SPECTRA ON VIEWING 
GEOMETRY}

\author{ A. WO\'ZNA, J. DYKS, T. BULIK \& B. RUDAK}

\address{Nicolaus Copernicus Astronomical Center, Rabia\'nska 8 \\
Toru\'n 87-100, Poland}

\maketitle\abstracts{We calculate the gamma-ray lightcurves and spectra of
rotation powered pulsars for a wide range of observer's viewing angles~($\zeta$), 
and inclination angles of the magnetic axis to the rotation axis~($\alpha$). 
We show how the shapes of the observed phase-averaged spectra as well as
energy-integrated pulse profiles depend on the orientation for classical and
millisecond pulsars.}

We compute characteristics of high energy emission from pulsars within the 
framework of a polar cap model. We use the Monte Carlo code  
described in detail in papers~\cite{{dh},{dr}} to show how the shapes of 
the observed phase-averaged spectra and the corresponding pulse 
profiles depend on viewing geometry and pulsar parameters.
We assume that the magnetic field is a rigid dipole tilted at an angle 
$\alpha$ with respect to the rotation axis. 
In the calculation we include the effects of aberration and the time of 
flight delays (TOF). These effects are important when the site of gamma-ray
emission moves at relativistic speeds, i.e. for millisecond pulsars.
Figure~1 presents these two effects for a perpendicularly rotating 
millisecond pulsar with $P=2.3$\,ms and $B=10^{9}$\,G.
An observer is located at angle $\zeta \approx 90^{\circ}$ with respect to
the rotation axis. 

Figure~2 shows the characteristics of high-energy radiation calculated for a
classical pulsar with $P=0.1$\,s and $B=10^{12}$\,G 
for the magnetic dipole inclination $\alpha=60^{\circ}$. 
The top left panel (a) presents intensity distribution for 
radiation above 100~MeV as a function of 
the rotational phase~$\phi$ and the viewing angle~$\zeta$.
Positions of six observers located at different values of $\zeta$
are marked with horizontal lines of different type. Phase-averaged spectra 
recorded by each of these observers (panel b) and corresponding pulse profiles 
(panels c-h) are shown with the same line type as in the panel a. In panel
b, the spectrum of total emission from pulsar (i.e. integrated over all 
directions) is shown as a thin solid line. Its position relative to the 
phase-averaged observed spectra is arbitrary. 
Figure 3 is for the same $P$ and $B$ but for
$\alpha=8^{\circ}$ and a different choice of observer positions.
In Figures 4 and 5 we present the similar characteristics of
radiation for millisecond 
pulsar parameters ($P=2.3$\,ms and $B=10^{9}$\,G) for $\alpha=60^{\circ}$ 
and $\alpha=8^\circ$, respectively. For classical
pulsars (Figs 2 and 3), curvature as well as synchrotron emission 
is included, whereas for millisecond pulsars (Figs 1,4 and 5)
only the curvature radiation is calculated.

\section*{Conclusions}
\begin{itemize}
\item{The combined effects of aberration and retardation (TOF) lead to
spreading of the leading peak and to piling up of photons in the trailing
peak in the case of double-peak lightcurves (Fig 1, Fig 4 - panels e,f,g). 
The same mechanism leads to 
asymmetry of single peaks in pulsar lightcurves -- the leading wing is 
broader than the trailing one (Figs 2c,d,g, 4c,d,h). The latter effect is
noticeable even for the case of classical pulsars as long as the line of
sight misses the polar cap beam and the radiation in the peak originates at
high altitudes (Fig 2c,d,g).}
\item{In the case of highly inclined millisecond pulsars the position
of high energy 
cutoff is significanty different for the leading and the trailing peaks 
(for details see Dyks~\&~Rudak~\cite{dd}). This leads to step--like
decline above $10^5$ MeV (dashed, dot-dashed and three-dot-dashed 
lines in Figure 4b).}
\item{The energy spectra integrated over the entire sky (thin solid
line in panels b, Figs. 2-5) are different from the phase-averaged spectra 
for most viewing orientations.}
\end{itemize}

\begin{figure}[h]
\begin{center}
\includegraphics[scale = 0.57, angle= 270]{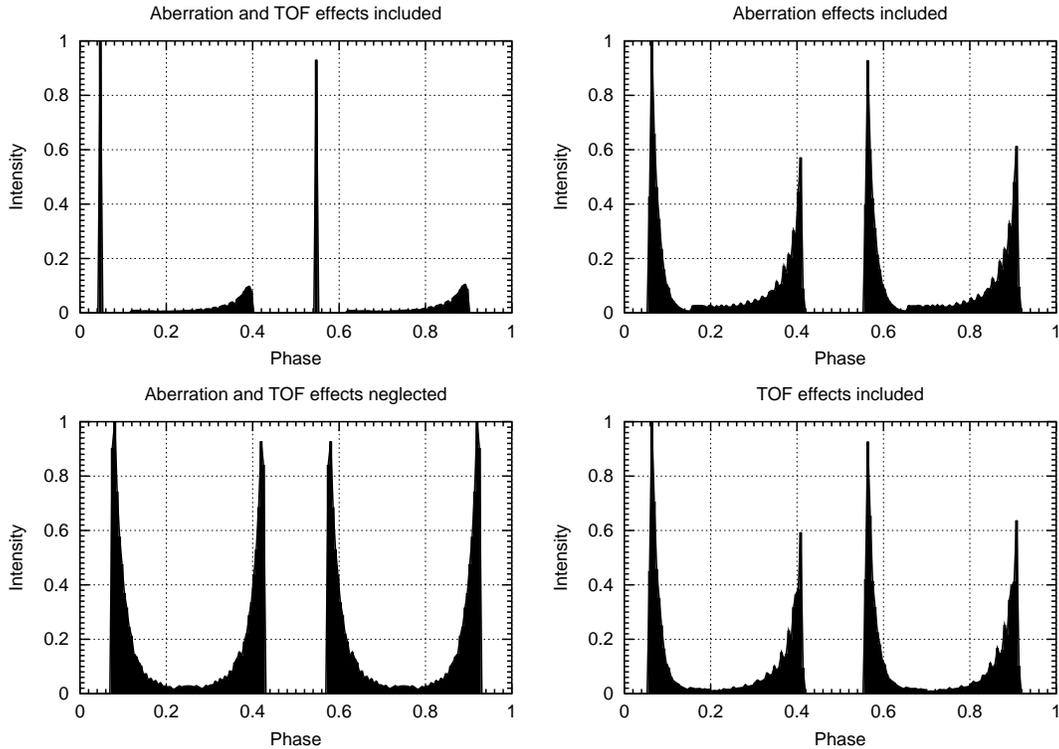}
\end{center}
\caption{The effects of frame transformation 
(photon aberration) and time of flight (TOF, retardation) for a 
millisecond pulsar ($P=2.3$\,ms, $B=10^{9}$\,G, $\alpha=90^{\circ}$, 
$\zeta \approx 90^{\circ}$). Four pulse shapes in the range 
above $100$\,MeV are shown. The upper left one is the result of 
calculations with both effects taken into account. To demonstrate their
importance we present how the pulse shapes look like when the effects are
ignored: the bottom left one corresponds to the case 
where we neglect the effects of aberration and retardation; the upper right,
and the bottom right pulse shapes correspond to the cases where we turn on 
the effects of aberration, and retardation, respectively.}
\end{figure}

\begin{figure}[h]
\begin{center}
\includegraphics[scale=0.895]{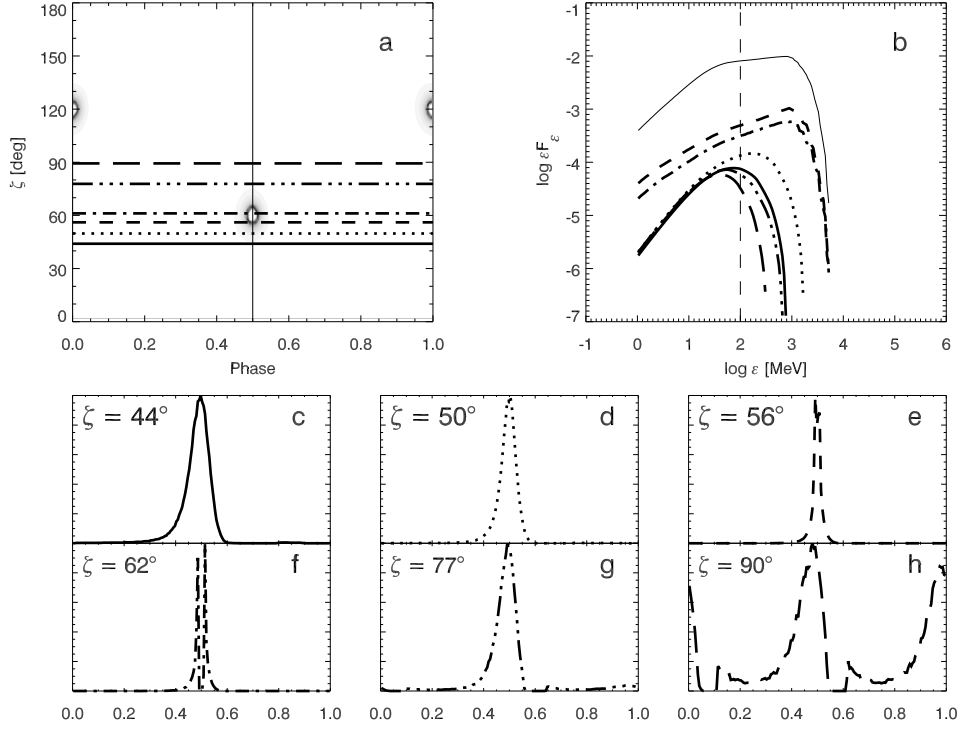}
\end{center}
\caption{The characteristic of the high energy radiation from a classical
pulsar with $P=0.1$\,s, $B=10^{12}$\,G and the magnetic dipole inclination  
$\alpha = 60^{\circ}$. The top left panel shows the intensity distribution
for radiation above $100$\,MeV as a function of rotational phase~$\phi$ 
and viewing angle $\zeta$. Horizontal lines  correspond to the
position of six observers located at different values of $\zeta$ (marked
with different line-type). For each of these observers the phase-averaged
spectra are shown (panel {\bf b}) with corresponding pulse profiles 
(panels {\bf c-h}, respectively) with the same line-type as in panel {\bf a}. 
The spectrum of total emission from the pulsar (thin
solid line in panel {\bf b}) is plotted at arbitrary level relative to the 
phase-averaged spectra.}
\end{figure}

\begin{figure}[h]
\begin{center}
\includegraphics[scale=0.895]{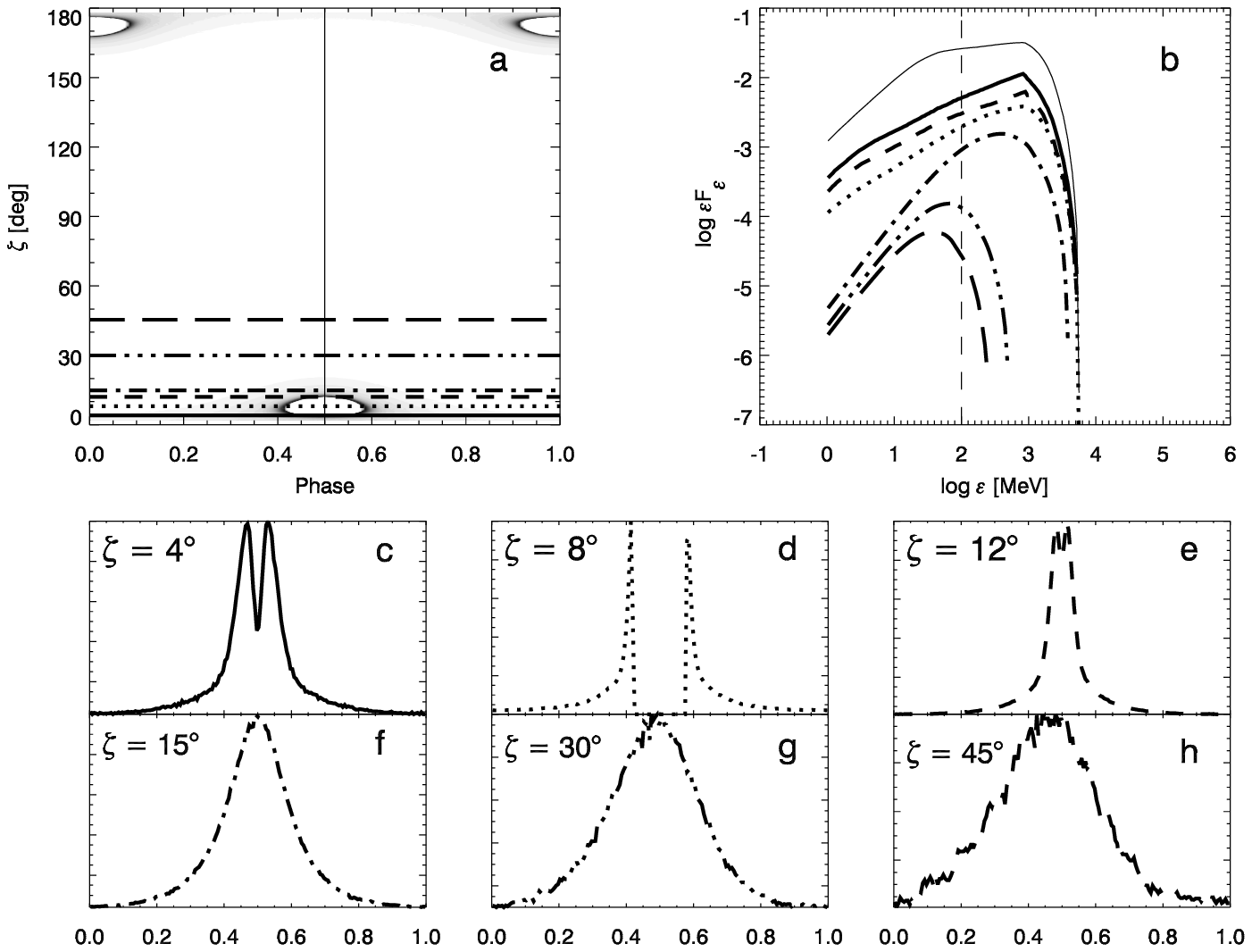}
\end{center}
\caption{Pulsar parameters are the same as in Figure 2, but 
for the inclination angle $\alpha = 8^{\circ}$, and a different set 
of $\zeta$ values.}
\end{figure}

\begin{figure}[h]
\begin{center}
\includegraphics[scale=0.9]{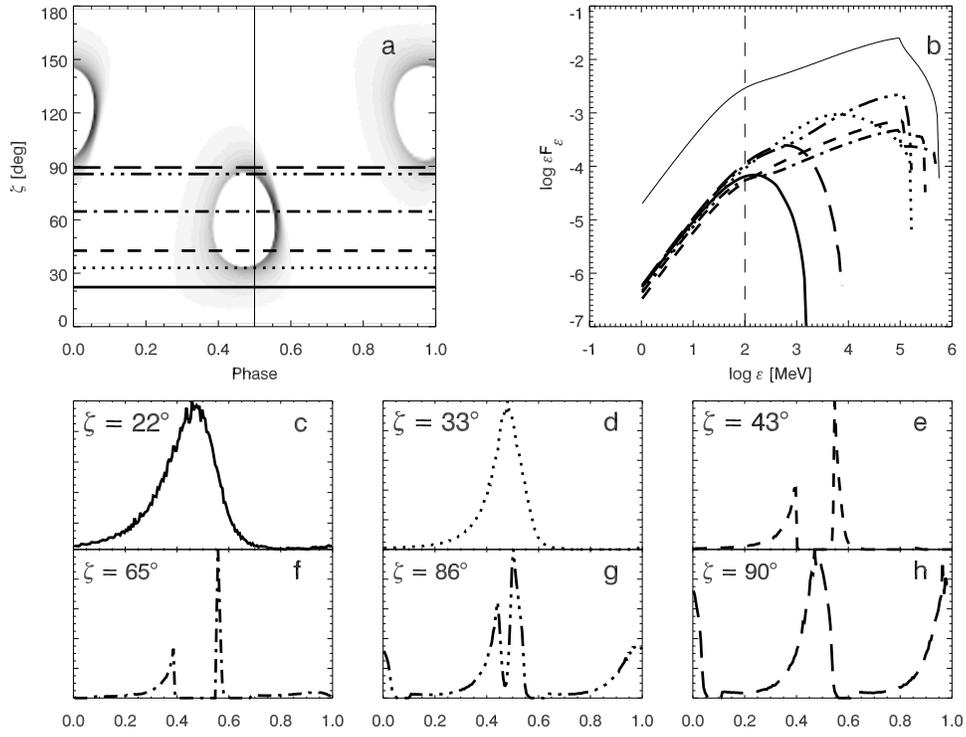}
\end{center}
\caption{Same as Figure 2, $\alpha = 60^{\circ}$, but for millisecond pulsar 
($P=2.3$\,s, $B=10^{9}$\,G) and for different viewing angles $\zeta$.}
\end{figure}

\begin{figure}[h]
\begin{center}
\includegraphics[scale=0.9]{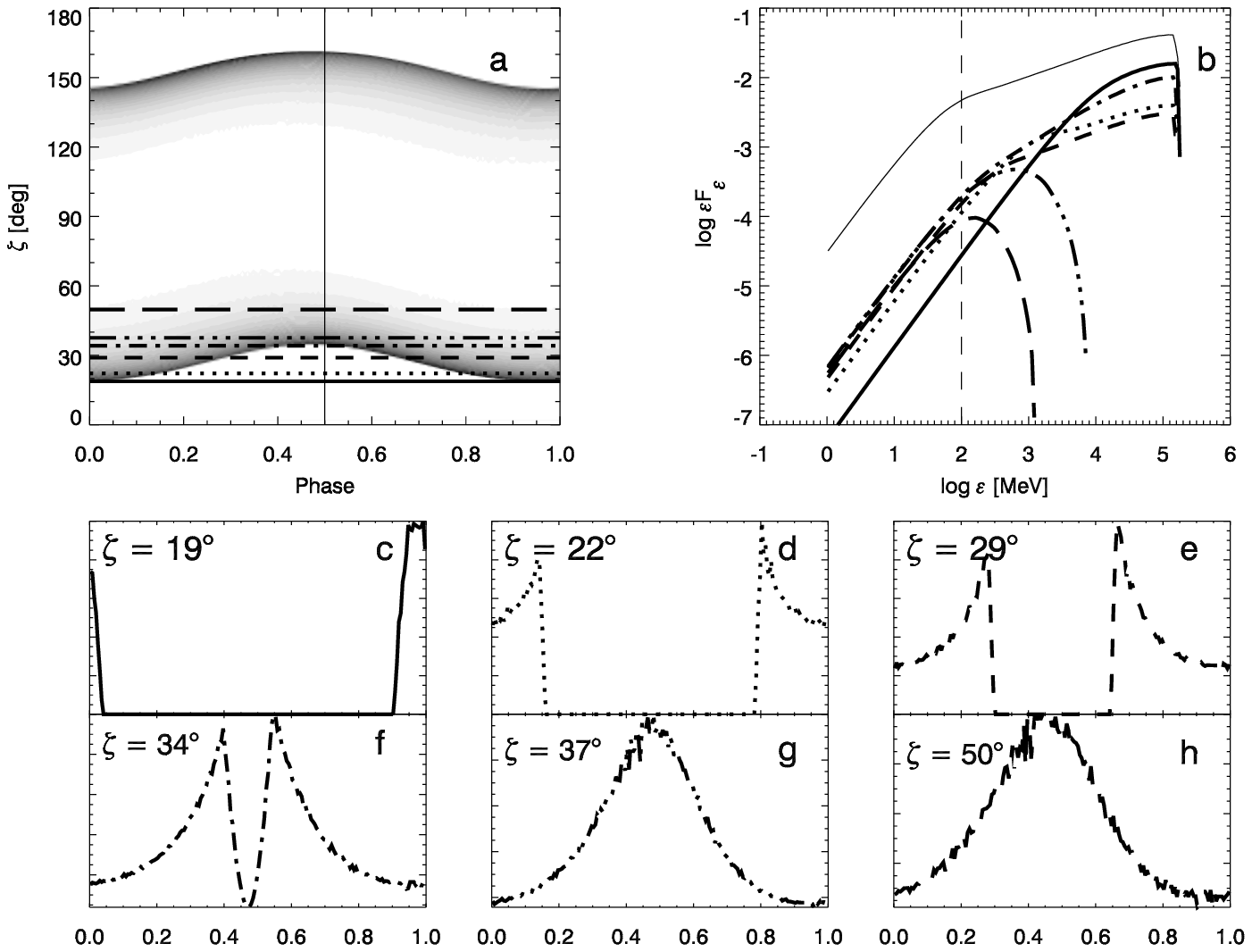}
\end{center}
\caption{Pulsar parameters are the same as in Figure 4, but 
for the inclination angle $\alpha = 8^{\circ}$, and
a different set of $\zeta$ values.}
\end{figure}

\section*{Acknowledgements}
This work was supported by the KBN grant 2P03D02117.


\section*{References}

\begin{thebibliography}{99}

\bibitem{dh}J.K. Daugherty and A. Harding, \Journal{\ApJ}{458}{278}{1996}.

\bibitem{dr}J. Dyks and B. Rudak, \Journal{\MNRAS}{319}{477}{2000}.

\bibitem{dd}J. Dyks and B. Rudak, {\em A\&A}, submitted

\end{thebibliography}
\end{document}